\documentstyle[12pt,a4wide]{article}

\newcommand{\be}[1]{\begin{equation}\label{#1}}
\newcommand{\ee}{\end{equation}}
\newcommand{\ba}[1]{\begin{eqnarray}\label{#1}}
\newcommand{\ea}{\end{eqnarray}}
\begin{document}
\hfill    {\bf FUB - HEP/94 - 1}

%\noindent
%{\small file: thetaneg.tex\\
%Version from \today}

\bigskip

\bigskip

%*****************************************************************
%*****************************************************************

\centerline{\LARGE\bf CLASSICAL AND QUANTUM }
\bigskip
\centerline{\LARGE\bf  BEHAVIOUR OF MULTIDIMENSIONAL}
\bigskip
\centerline{\LARGE \bf   INTEGRABLE COSMOLOGIES}
%%%%%%%%%%%%%%%%%%%%%%%%%%%%%%%%%%%%%%%%%%%%%%%%%%%%%%%%%%%%%%%

\bigskip

\bigskip
\centerline{ \bf U. BLEYER
\footnote{This work was supported by   WIP grant 016659}
}
\centerline{WIP-Gravitationsprojekt, Universit\"at Potsdam }
\centerline{An der Sternwarte 16 }
\centerline{ D-14482 Potsdam, Germany}
\bigskip
\centerline{and}
\bigskip
\centerline{\bf A. ZHUK
\footnote{Permanent address: Department of Physics,
University of Odessa,
2 Petra Velikogo,
Odessa 270100, Ukraine}
\footnote{This work was supported in part by DAAD and by DFG grant 436
UKR - 17/7/93}
}
\centerline{WIP-Gravitationsprojekt, Universit\"at Potsdam }
\centerline{An der Sternwarte 16 }
\centerline{ D-14482 Potsdam, Germany}
\centerline{and}
\centerline{Fachbereich Physik, Freie Universit\"at Berlin}
\centerline{Arnimallee 14}
\centerline{ D-14195 , Germany}

%*****************************************************************
%*****************************************************************

\begin{abstract}

Multidimensional cosmological models with $n~(n > 1)$ Einstein spaces
are discussed classically and with respect to canonical quantization.
These models are integrable in the case of  Ricci flat internal
spaces. For negative curvature of the external space we find exact
classical solutions modelling dynamical as well as spontaneous
compactification of the internal spaces. Spontaneous compactification
turns out to be an attractor solution. Solutions of the quantum
Wheeler-DeWitt equation are also obtained. Some of them describe the
tunneling process to be interpreted as the birth of the  universe from
''nothing''.

\end{abstract}

%*****************************************************************
%*****************************************************************

\newpage

\section{INTRODUCTION}

\noindent
Everyday experience seems to show evidently that we are living in a
four dimensional space-time. Why should we speak about a
multidimensional universe? There are good reasons to do so. First of
all, we know that consistent theories unifying fundamental
interactions take place in multidimensional spaces only, and may be
this is not purely a question of mathematical methods and extra
dimensions are a physical reality. Second, extra dimensions are
actually not observable, because they are extremely small at present
time. If they are of the scale of Planck's  length ($L_{PL} \sim
10^{-33}$cm) their observation is impossible due to the super high
frequency (energy) necessary to observe internal dimensions, and
behind Planck's lenght quantum uncertainties forbid the observation.
Nevertheless, during the evolution of the universe all dimensions
including ours and internal ones might have been of the same scale at
early stages of the universe. Moreover, extra dimensions could be much
larger then ours at that time. Thus, there is a reason to investigate
multidimensional cosmological models and the observable consequences
of the possible existence of extra dimensions.

In all multidimensional cosmological models (MCM) a mechanism of
dimensional reduction or, in other words, compactification of the
extra dimensions should be present. There are two approaches to
realize compactification. In the first case the internal dimensions
become much smaller than our external ones during the evolution of the
universe. These are the MCM with {\em dynamical compactification}.
Observable consequences of extra dimensions are in this case possible
variations of effective constants of nature (like the gravitation
constant)\cite{1} - \cite{4}, imprints in cosmic rays of ultrahigh
energy \cite{5} or in the spectrum of gravitational waves \cite{6}.
Another possibility consists in the proposal that all extra dimensions
are static and small from the very beginning. Such MCM are called
models with {\em spontaneous compactification}. The presence  of extra
dimensions leads in this case to the generation of  particle masses
\cite{7} - \cite{9}. In both of these approaches the presence of
extra dimensions has very strong influence on the evolution of our
external space-time. Compactification takes place for pure gravity as
well as for gravity coupled to different matter fields. There is a
large amount of papers devoted to these questions (see e.g. the
references in \cite{10}).

In our paper we shall consider MCM which consist of $n~(n > 1)$ spaces
of constant curvature. This model was investigated from the classical
as well as from  quantum points of view in papers \cite{11} -
\cite{20}. The model can be generalized to the case of all spaces
being Einstein spaces. The multidimensional Einstein equations as well
as the quantum Wheeler-DeWitt equation (WDW) can be integrated for
this model if at most one of the spaces of constant curvature $M_i~(i
= 1, \dots, n)$ is not Ricci flat \cite{11}, \cite{14}, \cite{17}.
This property is not changed if the model contains in addition a
massless minimally coupled scalar field. From the point of view of
dynamical and spontaneous compactification this integrable model was
considered in \cite{10} in the case when the non Ricci flat space is
of positive constant curvature. This space, let it be $M_1$, was
considered there as our external space and all other factor spaces are
internal ones. Both types of compactification were found. In the case
of a real scalar field as matter source in the Lorentzian region the
solution with spontaneous compactification permits an intersting
continuation to the Euclidean region describing Euclidean wormholes.

In the present paper we shall investigate this integrable models
on the classical as well as on the quantum levels for the case when
the non Ricci flat space $M_1$ is of constant negative curvature. The
main problem consists in the investigation of the dynamical and
spontaneous compactification. In the case of positive curvature of
$M_1$ the parameter playing the role of energy may  in the
Lorentzian region adopt  positive values only \cite{14}. In contrast to this
case, the model with negative curvature permits positive as well as
non positive values of this parameter. This feature of the models with
negative curvature leads to a more rich and interesting picture than
in the former case.

It will be shown that the MCM investigated here have solutions
describing spontaneous and dynamical compactification. The paper is
organized as follows. In Section 2. we describe our MCM and represent
the classical Einstein equations for this model in appropriate
coordinates. Section 3. is devoted to the investigation of the
dynamical as well as spontaneous compacdtification on the classical
level. In Section 4. we consider the quantum properties of the model.
The results of the paper are summarized in the Conclusions. A
particular case of the model with dynamical compactification is
presented in an Appendix. References complete the paper.

%*****************************************************************
%*****************************************************************

\section{MINISUPERSPACE  COSMOLOGICAL MODELS}

%*****************************************************************

Let us consider the metric
\be{1}
ds^{2} = - d\tau^{2} exp(2\gamma (\tau)) + \sum_{i=1}^{n}
e^{2\beta^{i}(\tau)}  ds_{i}^{2}
\end{equation}
on a $D$-dimensional space-time manifold
\be{2}
M = R \times M_{1} \times \dots \times M_{n}
\end{equation}
where the $M_{i}$ are  $d_{i}$-dimensional compact spaces of constant
curvature with line elements $ ds_{i}^{2}$. The connection to the
scale factors $a_{i}$ is given by $a_{i} = e^{\beta^{i}}$.
The scalar curvature of $M_{i}$ is normalized in such a way that we
can write
\be{3}
R[g_{(i)}] = \theta_{i} = k_{i}d_{i}(d_{i} - 1), \qquad i=1, \dots, n,
\ee
where $k_{i} = 0, \pm 1$.
In the case of non positive curvature  the compactness
condition for the internal spaces can be achieved by appropriate
periodicity conditions for the coordinates \cite{21}.

As mentioned in the Introduction, this model can be  generalized to
the case of Einstein spaces $M_i$ for which $R[g_{(i)}] =
\lambda_id_i$ instead of (\ref{3}) and $\lambda_i$ are arbitrary
numbers. In formulas obtained later on this generalization is achieved
by the trivial substitution $\theta_i \rightarrow \lambda_id_i$.

We restrict our consideration to the important case when only one of
the spaces $M_{i}$ is not Ricci flat with negative curvature:
$\theta_{1} < 0, ~\theta_{i} = 0, ~i = 2, \dots , n$. In this case the
cosmological model is a completely integrable system \cite{14,17}.
This can be generalized by taking into account a minimally coupled
free scalar field $\varphi$.

The  action $S$ for the model with the metric
(\ref{1}) and a minimally coupled scalar field can be written
in the form \cite{11}
\be{4}
S = \int {\cal L} d t
\end{equation}
where the Lagrangian has the form
\ba{5}
{\cal L} = \frac{\mu}{2}e^{-\gamma + \sum_{i=1}^{n} d_{i}\beta^{i} }
\left \{\sum_{j=1}^{n} d_{j} (\dot\beta^{j})^{2} - (\sum_{j=1}^{n} d_{j}
\dot\beta^{j})^{2} + \kappa^{2}\dot\varphi^{2}\right \}  \nonumber \\
- \frac{\mu}{2}  e^{\gamma + \sum_{i=1}^{n} d_{i}\beta^{i}}
\mid \theta_{1} \mid e^{-2\beta^{1}}
\ea
Here $\kappa^2$ denotes the gravitational constant and $\mu =
\prod_{i=1}^{n} V_{i}/\kappa^{2}$ where $V_{i}$ is the volume of
$M_{i}$: $V_{i} = \int_{M_{i}} d^{d_{i}} y
(det(g_{m_{i}n_{i}}))^{1/2}$. The metric (\ref{1}) can be normalized
in such a way that $\mu = 1$. After normalizing $\mu$ we may also
consider noncompact spaces. We shall also use natural units with
$\kappa^2 = 1$.

The analysis of this system will be done mostly in two time gauges,
in the gauge of harmonic time $\tau$
%%%%%%%%%%%%%%%%%%%%
\cite{11}
%%%%%%%%%%%%%%%%%%%%%
where $\gamma = \sum_{i=1}^{n} d_{i}\beta^{i}$, and in the gauge of
synchronous time $t$ with $\gamma = 0$.

The possibility of the free choice of gauge implies the following
constraint equation
\be{6}
e^{-\gamma}
\left[ \sum_{i=1}^{n} d_{i} (\dot\beta^{i})^{2} - (\sum_{i=1}^{n} d_{i}
\dot\beta^{i})^{2} + \dot\varphi^{2} \right]
+ e^{- \gamma}\mid \theta_{1} \mid e^{-2\beta^{1}} = 0
\ee
It was shown in
%%%%%%%%%%%%%%%%%%%%%%%%%%%%%%%%%%%%%%%%%%%
\cite{14}
%%%%%%%%%%%%%%%%%%%%%%%%%%%%%%%%%%%%%%%%%%%%%%
that the field equations for this model can be integrated most easily
using the following coordinates:
\ba{7}
q v^{0} & = & (d_{1} - 1) \beta^{1} + \sum_{i=2}^{n} d_{i}\beta^{i},
                   \nonumber \\
q v^{1} & = & \left[(D -2)/(d_{1} {\sum}_{2} \right]^{\frac{1}{2}}
             \sum_{i=2}^{n} d_{i}\beta^{i},  \\
q v^{i} & = & \left[(d_{1} -1)d_{i}/(d_{1} {\sum}_{i}{\sum}_{i+1}
            \right]^{\frac{1}{2}}
             \sum_{j=i+1}^{n} d_{j}(\beta^{j} - \beta^{i}),
          \hspace{1cm} i = 2, \dots, n-1 \nonumber
\end{eqnarray}
Here we used the notation $D = 1 + \sum_{i=1}^n d_i, ~q^{2} =
(d_{1} - 1)/d_{1}$, and $\sum_i = \sum_{j=i}^n d_{j}$.
In the Lorentzian region and in harmonic time gauge the Lagrangian and
the constraint equation take the form
\be{8}
{\cal L} = \frac{1}{2} \left( - (\dot v^{0})^{2} + \sum_{i=1}^{n-1}
(\dot v^{i})^{2} + \dot\varphi^{2} \right) - \frac12 \mid \theta_{1} \mid
e^{2q v^{0}}
\ee
and
\be{9}
- (\dot v^{0})^{2} + \sum_{i=1}^{n-1}
(\dot v^{i})^{2} + \dot\varphi^{2} + \mid \theta_{1} \mid
e^{2q v^{0}} = 0
\ee
The dot denotes the derivative with respect to the harmonic time
$\tau$. The consideration of the system can be generalized to the
Euclidean region by analytic continuation.

The equations of motion corresponding to the Langrangian (\ref{8})
read
\ba{10}
\ddot v^{0} - q \mid \theta_{1} \mid e^{2q v^{0}} & = & 0 \nonumber \\
\ddot v^{i} & = & 0, \hspace{1cm} i = 1, \dots , n - 1 \\
\ddot\varphi & = & 0 \nonumber
\ea
The last two equations are easy to integrate. We find
\ba{11}
v^{i} & = &\nu^{i}\tau + c^{i}, \hspace{1cm} i = 1, \dots , n-1
\nonumber \\
\varphi  & = & \nu^{n}\tau + c^{n}
\ea
where the $\nu^{i}$ and $c^{i}$ are constants of integration.
In minisuperspace of vectors $\vec v = (v^0, v^1, \dots,
v^{n-1},v^n\equiv \varphi)$ the indices are raised and lowered by the
diagonal metric $\eta = (-1, +1, \dots,+1)$ \cite{14}. Thus, we have
$v^0=-v_0, v^i=v_i, \nu^i=\nu_i$ and $c^i=c_i, i=1, \dots, n$.
Now the
constraint equation may be rewritten
\be{12}
(\dot v^{0})^{2} - |\theta_{1}| e^{2q v^{0}} = \epsilon
\ee
with
\be{13}
\epsilon = \sum_{i = 1}^{n} (\nu^{i})^{2}
\ee
It can be seen from (\ref{12}) that $\epsilon$ can be treated as an
energy. This was shown in more detail in
%%%%%%%%%%%%%%%%%%%%%%%%5
\cite{14}.
%%%%%%%%%%%%%%%%%%%%%%%%%%

%%%%%%%%%%%%%%%%%%%%%%%%%%%%%%%%%%%%%%%%%%%%%%%%%%%%%%%%%%%%%%%%%%%%%%
\section{CASSICAL SOLUTIONS}
%%%%%%%%%%%%%%%%%%%%%%%%%%%%%%%%%%%%%%%%%%%%%%%%%%%%%%%%%%%%%%%%%%%%%%

Equations (\ref{10} - \ref{13}) are written in the Lorentzian region.
The parameters $\nu_i ~(i = 1, \dots, n)$ are the momenta in
minisuperspace. Thus, $E \equiv \frac12 \epsilon$ plays the role of
energy \cite{14}. Equation (\ref{12}) shows that we can consider the
case $\epsilon \le 0$ as well as $\epsilon > 0$. We have to demand the
metric to be real in the Lorentzian region. In what follows, the
momenta $\nu_i~(i = 1, \dots, n-1)$ should be real there (see eqn.
(\ref{11}). The case $\epsilon = 0$ is treated as ground state where
all momenta are putted equal to zero: $\nu_i = 0 ~(i = 1, \dots, n)$.
For $\epsilon > 0$ all $\nu_i  ~(i = 1, \dots, n)$ are considered to
be arbitrary real numbers. The case $\epsilon < 0$
is a little bit more complicated. Here the demand of a real metric in
the Lorentzian region leads to the condition that all $\nu_{i} ~(i
= 1, \dots, n - 1)$ are real and the condition $\epsilon < 0$ and eqn.
(\ref{13}) are compatible only for a purely imaginary $\nu_{n}$. This
means  that the scalar field has in this case to be imaginary in the
Lorentzian region and we have the following condition
\be{14}
\sum_{i= 1}^{n - 1}(\nu_{i})^{2} - \mid \nu_{n} \mid^{2} < 0
\ee
Let us consider the three special cases $\epsilon = 0$, $\epsilon < 0$,
and $\epsilon > 0$ separately.

%%%%%%%%%%%%%%%%%%%%%%%%%%%%%%%%%%%%%%%%%%%%%%%%%%%%%%%%%%%%%%%%%%%%%%%
\subsection{The case $\epsilon = 0$}
%%%%%%%%%%%%%%%%%%%%%%%%%%%%%%%%%%%%%%%%%%%%%%%%%%%%%%%%%%%%%%%%%%%%%%%
%%%%%%%%%%%%%%%%%%%%%%%%%%%%%%%%%%%%%%%%%%%%%%%%%%%%%%%%%%%%%%%%%%%%%%%

As shown above in this case we have $\nu_{i} = 0 ~(n = 1, \dots, n)$.
Then it can be seen from eqns. (\ref{11}) and (\ref{7}) that all scale
factors are frozen ($a_{i}= e^{\beta^{i}} = a_{0(i)}, ~i = 2, \dots, n,
solutions we call {\it solutions with spontaneous compactification}.
In this case we have only one scale factor with dynamical behaviour
(in our case $a_{1}$) and the corresponding factor space $M_{1}$ will
be associated with the external (our real) space. The fixed scale
factors are assumed to be of the order of Planck length $a_{(0)i} \sim
L_{Pl} \sim 10^{-33}$cm $(i = 2, \dots, n)$ and the
associated factor spaces are looked at as unobservalbe internal
dimensions.

The dynamical behaviour of the remaining developing scale factor can
be defined from the solution of eqn. (\ref{12})
\be{15}
e^{qv^{0}} = \left( (d_{1} - 1)\mid \tau \mid \right)^{-1},
\hspace{1cm} - \infty < \tau < \infty
\ee
With the help of transformation (\ref{7}) we find the expression
for the scale factor $a_{1}$ in hamonic time gauge
\be{16}
a_{1}(\tau) = \left[ 1/C(d_{1} - 1)\mid\tau\mid   \right]^{1/(d_{1} - 1)}
\ee
where
\be{17}
C = \prod_{i=2}^{n} a_{(0)i}^{d_{i}}
\ee
Then the metric in harmonic time takes the form
\be{18}
g = -e^{2\gamma(\tau)} d\tau\otimes d\tau + a_{1}^{2}(\tau) g_{(1)} +
\sum_{i = 2}^{n} a_{(0)i}^{2} g_{(i)}
\ee
where
\be{19}
e^{\gamma(\tau)} = C a_{1}^{d_{1}}(\tau)
\ee
It is worth to rewrite the metric in synchronous time coordinate $t$.
In this case we have $a_{1}(t) = \mid t \mid$ and the metric is given
by
\be{20}
g = - dt\otimes dt + t^{2} g_{(1)} +
\sum_{i = 2}^{n} a_{(0)i}^{2} g_{(i)}
\ee
{}From this expression we see that the dynamical part of the universe is
described the Milne model.
%%%%%%%%%%%%%%%%%
[\cite{22}].
%%%%%%%%%%%%%%%%%%%%%%%%%%%%%%%%%%%%%%%%%
%\marginpar{Def. Milne universe}
%%%%%%%%%%%%%%%%%%%%%%%%%%%%%%%%%%%%%%%%%
In this way we find for the physically interesting case of
Kaluza-Klein theory with $d_{1} = 3$ and $M_1$ being an open
hyperbolic space that the solution with $\epsilon
= 0$ describes the following topology of a spontaneously compactified
universe
\be{21}
M^{4} \times T^{d_{2}} \times \dots \times T^{d_{n}}
\ee
Here $M^{4}$ is the four dimensional Minkowski space-time and the
$T^{d_{i}}$ are $d_{i}$- dimensional Tori (or other compact spaces of
constant zero curvature). The Tori are frozen and asumed to have
scales of Planck size.

%%%%%%%%%%%%%%%%%%%%%%%%%%%%%%%%%%%%%%%%%%%%%%%%%%%%%%%%%%%%%%%%%%%%%%%
\subsection{The case $\epsilon > 0$}
%%%%%%%%%%%%%%%%%%%%%%%%%%%%%%%%%%%%%%%%%%%%%%%%%%%%%%%%%%%%%%%%%%%%%%%
%%%%%%%%%%%%%%%%%%%%%%%%%%%%%%%%%%%%%%%%%%%%%%%%%%%%%%%%%%%%%%%%%%%%%%%

As explained above, in this case all parameters $\nu_{i}, (i = 1,
\dots, n)$ are considered to be real.
The solution of eqn. (\ref{12}) takes the form
\be{22}
e^{qv^{0}} = \frac{\sqrt{\epsilon/\mid \theta_{1} \mid}}{\mid
\sinh\left[ (d_{1} - 1)\sqrt{\epsilon/\mid \theta_{1} \mid}(\tau -
\tau_{0}) \right] \mid}, \hspace{0.5cm} - \infty < \tau < + \infty
\ee
where $\tau_{0}$ is a constant of integration. Choosing the initial
value of the harmonic time coordinate in an appropriate way we can put
$\tau_{0} = 0$.

The eqns. (\ref{22}) and (\ref{11}) give a general solution of the
system (\ref{10}) with constraint (\ref{12}).
But in cosmology usually the synchronous time
coordinate $t$ is used and it is quite difficult to find the
dependence of the scale factors $a_{i} = e^{\beta^{i}}$ on this time
coordinate. But the explicit dependence on $t$ can be found in some
interesting special cases which we now consider.

%%%%%%%%%%%%%%%%%%%%%%%%%%%%%%%%%%%%%%%%%%%%%%%%%%%%%%%%%%%%%%%%%%%%
\subsubsection{The 2-component universe. Dynamical compactification}
%%%%%%%%%%%%%%%%%%%%%%%%%%%%%%%%%%%%%%%%%%%%%%%%%%%%%%%%%%%%%%%%%%%%
%%%%%%%%%%%%%%%%%%%%%%%%%%%%%%%%%%%%%%%%%%%%%%%%%%%%%%%%%%%%%%%%%%%%

Let us consider the special case where only two factor spaces are
included in the model. We shall show that in this case solutions with
dynamical compactification occur, that means solutions with both
scale factors depending on time but with one, let it be $a_{1}$, increasing and
the other one ($a_{2}$) shrinking to Planck scales. In this case
$M_{1}$ is treated as our external space and $M_{2}$ describes an unobservable
internal space.
For two component cosmological models (i.e. for $n = 2$ in eqns.
(\ref{11}) and (\ref{22})) it is easy to get the explicit expressions
for the scale factors as functions of harmonic time:
\be{23}
a_{1}^{d_{1} - 1} = \frac{2 {a_{(0)}}_{1}^{d_{1} - 1}}
{e^{\sqrt{\frac{d_{2}(d_{1} - 1)}{D - 2}}~\nu_{1}\tau}
\left[ e^{\sqrt{\frac{(d_{1} - 1)(\nu_{1}^{2} +
\nu_{2}^{2})}{d_{1}}}~|\tau |}
  -    e^{- \sqrt{\frac{(d_{1} - 1)(\nu_{1}^{2} +
\nu_{2}^{2})}{d_{1}}}~|\tau |}
\right]}
\ee
\be{24}
a_{2}^{d_{2}} = {a_{(0)}}_{2}^{d_{2}} e^{\sqrt{\frac{d_{2}(d_{1} -
1)}{D - 2}}~\nu_{1}\tau}
\ee
Here ${a_{(0)}}_{1} $ and ${a_{(0)}}_{2} $ are connected with the
constant of integration $c_{1}$ in (\ref{11}) by the expressions
\be{25}
{a_{(0)}}_{1}^{d_{1} - 1} = \sqrt{\frac{\nu_{1}^{2} +
\nu_{2}^{2}}{d_{1}(d_{1} - 1)}} e^{- \sqrt{\frac{d_{2}(d_{1} -
1)}{D - 2}}~c_{1}}
\ee
\be{26}
{a_{(0)}}_{2}^{d_{2}} =  e^{\sqrt{\frac{d_{2}(d_{1} -
1)}{D - 2}}~c_{1}}
\ee
{}From this we get the connection between ${a_{(0)}}_{1}$ and
${a_{(0)}}_{2}$
\be{27}
{a_{(0)}}_{1}^{d_{1} - 1}{a_{(0)}}_{2}^{d_{2}} = \sqrt{\frac{\nu_{1}^{2} +
\nu_{2}^{2}}{d_{1}(d_{1} - 1)}}
\ee
It can be seen from (\ref{23}) that $a_{1}$ has a discontinuity  at
$\tau = 0$. Therefore, we have different solutions in the region $-
\infty< \tau \le 0$ and $0 \le \tau < \infty$.

It follows from eqn. (\ref{23}) that we have three different types of
development of the scale factors $a_{1}$ and $a_{2}$ in dependence
{}from the sign of $\sqrt{d_{1}d_{2}\nu_{1}^{2}}  -
\sqrt{(d_{1} + d_{2} - 1)(\nu_{1}^{2} + \nu_{2}^{2})}$. Let us
consider these cases in more detail.

\bigskip
\noindent
1. $\hspace{1cm}  \sqrt{d_{1}d_{2}\nu_{1}^{2}} - \sqrt{(d_{1} + d_{2}
- 1)(\nu_{1}^{2} + \nu_{2}^{2})} > 0$
\bigskip

Here, in dependence on the sign of $\nu_{1}$ there  two
types of solutions exist for the scale factors (see qualitative pictures
fig. 1 and fig. 2). From
fig. 1 ($\nu_{1} > 0$) we see that for $\tau > 0$ dynamical
compactification occurs when $\tau \rightarrow + \infty$. For $\tau <
0$ the dynamical behaviour of the scale factors is more complicated.
Nevertheless, there are also time intervals where one of the scale
factors is much bigger than the other one, for example in the limit
$\tau \rightarrow -0$. For the case of fig. 2 ($\nu_{1} < 0$)
dynamical compactification occurs for $\tau < 0$ when $\tau
\rightarrow - 0$ or for $\tau > 0$ if $\tau \rightarrow + \infty$.
Moreover, depending
on the sign of $\nu_{1}$ the space $M_{1}$ or the space
$M_{2}$ can play the role of the external space. We have also to
mention that the case 1. can be realized if no scalar field is present
($\nu_{2} = 0$). It is clear that the condition 1. is not valid for
$d_2=1$.

\noindent
2. $\hspace{1cm}  \sqrt{d_{1}d_{2}\nu_{1}^{2}} - \sqrt{(d_{1} + d_{2}
- 1)(\nu_{1}^{2} + \nu_{2}^{2})} < 0$
\bigskip

The qualitative behaviour of the scale factors is shown in figs. 3, 4.
It can be seen that there are also in this case  regions with dynamical
compactification. Because of $d_{1}, d_{2} \ge 1$ the case 2. can not
be realized if $\nu_{2} = 0$, i.e. without a scalar field being
present.

\noindent
3. $\hspace{1cm}  \sqrt{d_{1}d_{2}\nu_{1}^{2}} - \sqrt{(d_{1} + d_{2}
- 1)(\nu_{1}^{2} + \nu_{2}^{2})} = 0$
\bigskip

In this case we have a connection between the two constants of
integration $\nu_{1}$ and $\nu_{2}$:
\be{28}
\nu_{2}^{2} = \left( \frac{d_{1}d_{2}}{d_{1} + d_{2} - 1} - 1
\right) \nu_{1}^{2}
\ee
{}From this expression we see that in the case $d_{2} = 1$ there is no
scalar field: $\nu_{2} = 0$. Using (\ref{28}), eqn. (\ref{27})
reads
\be{29}
a_{(0)1}^{d_{1} - 1}a_{(0)2}^{d_{2}} =
\sqrt{\frac{d_{2}}{(d_{1} - 1)(d_{1} + d_{2} - 1)}} \mid \nu_{1} \mid
\ee
In this case we find for the scale factors the following expressions
\be{30}
a_{1}^{d_{1} - 1} = \frac{2a_{(0)1}^{d_{1} - 1}}{\mid e^{\pm 2
\sqrt{\frac{d_{2}(d_{1} - 1)}{D - 2}}~\mid \nu_{1} \mid \tau } - 1 \mid}
\ee

\be{31}
a_{2}^{d_{2}} = a_{(0)2}^{d_{2}} \exp\left\{\sqrt{\frac{d_{2}(d_{1} - 1)}
{D - 2}}~\nu_{1} \tau \right\}
\ee
where the upper sign in the expression for $a_{1}$ corresponds to
$\nu_{1} > 0$ and the lower one to $\nu_{1} < 0$.
The qualitative behaviour of the scale factors in the case
$2^{1/(d_1-1)}a_{(0)1} > a_{(0)2}$ is given in figs. 5, 6
showing the existence of  regions with dynamical compactification, too.

Because of the simple form of eqn. (\ref{30}) we can find the
explicit connection between the conformal time $\tau$ and the
synchronous time $t$. This connection is given by \cite{11}
\be{32}
dt = \pm e^{\gamma} d\tau = \pm a_{1}^{d_{1}} a_{2}^{d_{2}} d\tau
\ee
Putting (\ref{30}) and (\ref{31}) into (\ref{32}) we find
\be{33}
t = \pm \tilde c \int \frac{dy}{\mid y^{2} - 1
\mid^{\frac{d_{1}}{d_{1} - 1}}}  + \tilde{\tilde c}
\ee
where
\be{34}
y = e^{\pm {\sqrt{\frac{d_{2}(d_{1} - 1)}{D - 2}}
\ee
The upper sign corresponds to $\nu_{1} > 0$ the lower one to $\nu_{1}
< 0$. The constant $\tilde c$ is defined by
\be{35}
\tilde c = 2^{\frac{d_{1}}{d_{1} - 1}} \sqrt{\frac{d_{1} + d_{2} - 1}
{d_{2}(d_{1} - 1)}} \frac{1}{\mid \nu_{1} \mid} a_{(0)1}^{d_{1}}
a_{(0)2}^{d_{2}}
\ee
and the initial value of the synchronous time will be taken
so that $\tilde{\tilde c} = 0$.

In the case $d_{1} = 2, d_{2} \geq 1$ the integration of (\ref{34}) gives
\be{36}
t = \pm \frac{\tilde c}{2}
       \left\{
             \begin{array}{ll}
               {{ \frac{y}{1 - y^{2}}} + {\mbox{arcoth} y}}, & \mbox{$ \mid y
                                                             \mid > 1 $}
                \\
                {{\frac{ y }{ 1 - y^{2}}} + {\mbox{artanh} y}}, & \mbox{$\mid
                                                           y \mid < 1 $}
             \end{array}
       \right.
\ee
This case is of special interest because for $d_{2} = 1$ it describes
a 3-dimensional anisotropic Bianchi III universe without scalar field
\cite{25}.
%%%%%%%%%%%%%%%%5
%[\cite{Louko}]
%%%%%%%%%%%%%%%%
As we could see above  there is no isotropization in this case.

The integral (\ref{33}) can be easyly calculated in the mostly
important for Kaluza-Klein theory case $d_{1} = 3$ ($M_{1}$ can be
looked at as our external space). Then we have two types of
solutions:
\ba{37} %{38}{39}
a_{1} & = & \frac{2^{1/2}a_{(0)1}}{\tilde c} (t^{2} - {\tilde
c}^{2})^{1/2}  \\
a_{2} & = & a_{(0)2} \left[ \frac{t^{2}}{t^{2} - {\tilde c}^{2}}
\right]^{\frac{1}{2d_{2}}} \\
\varphi & = & \pm \frac{1}{2} \sqrt{\frac{d_{2} - 1}{d_{2}}} \ln
\frac{t^{2}}{t^{2} - {\tilde c}^{2}}
\ea
where $\mid t \mid \geq \tilde c$
and
\ba{40} %{41}{42}
a_{1} & = & \frac{2^{1/2}a_{(0)1}}{\tilde c} (t^{2} + {\tilde
c}^{2})^{1/2}  \\
a_{2} & = & a_{(0)2} \left[ \frac{t^{2}}{t^{2} + {\tilde c}^{2}}
\right]^{\frac{1}{2d_{2}}} \\
\varphi & = & \pm \frac{1}{2} \sqrt{\frac{d_{2} - 1}{d_{2}}} \ln
\frac{t^{2}}{t^{2} + {\tilde c}^{2}}
\ea
where $ - \infty < t < + \infty$. In figures 7, 8 the behaviour of the
scale factors corresponding to (37), (38) and (40), (41)
is presented.
In both of these cases there are regions with  dynamical
compactification. For the solutions (37), (38) the scale factors
are everytimes in opposite phases and the role
of the exterior space can be played by $M_{1}$ as well as by $M_{2}$.
For instance, in the region $t \geq \tilde c$ the scale factor $a_{2}$
shrinks from $+ \infty$ to $a_{(0)2}$ (asymptotically). For $a_{(0)2}
\sim L_{Pl}$ $M_{2}$ becomes unobservable. At the same time $a_{1}
\sim t (t \gg \tilde c)$ and $M_{1}$ describes the exterior space of
Milne type. In this way the topology of the universe tends
asymptotically to $M^{4} \times T^{d_{2}}$.

Fig. 8 is drawn for the case $\sqrt{2} a_{(0)1} > a_{(0)2}$. For
$\mid t \mid \gg \tilde c$ we have $a_{2} \rightarrow a_{(0)2}$ and
$a_1 \sim |t|$, i.e.
the factor space $M_{2}$ becomes static and the factor space $M_1$ has
asymptotically the behaviour of a Milne universe. Thus, for $|t| \gg
\tilde c$ the topology of the universe tends asymptotically again to
$M^4\times T^{d_2}$.

Figures 9, 10 show the behaviour of the scalar field for the upper
sign in formulas (39) and (42) correspondingly. It is
interesting to note that $|\varphi| \rightarrow 0$ if $|t| \gg \tilde
c$.

Here, we have investigated the two-component model $(n=2)$ in the case
$\epsilon > 0$. There exists one particular case for $n > 2$ which may
be reduced to the two-component model considered above (s. Appendix).

%%%%%%%%%%%%%%%%%%%%%%%%%%%%%%%%%%%%%%%%%%%%%%%%%%%%%%%%%%%%%%%%%%%%%%%
\subsubsection{The n-component universe. Spontaneous compactification}
%%%%%%%%%%%%%%%%%%%%%%%%%%%%%%%%%%%%%%%%%%%%%%%%%%%%%%%%%%%%%%%%%%%%%%%
%%%%%%%%%%%%%%%%%%%%%%%%%%%%%%%%%%%%%%%%%%%%%%%%%%%%%%%%%%%%%%%%%%%%%%%

Like in the case $\epsilon = 0$ we have a solution with spontaneous
compactification also for $\epsilon > 0$. Obviously, this special case
corresponds to the following choice of contants of integration:
$\nu_{i} = 0 ~(i = 1, \dots, n-1), ~\nu_{n} \not= 0$. Therefore, this
case is only realized in the presence of a real scalar field. All scale
factors are frozen $(e^{\beta^{i}} = a_{i} = a_{(o)i}, i =2, \dots, n)$
except one $(e^{\beta^{1}} = a_{1})$. Then the metric in harmonic time
$\tau$ takes the form (\ref{18}) where the scale factor $a_{1}$ is
\be{43}
a_{1}(\tau) = \left( \sqrt{\epsilon / \mid \theta_{1} \mid}/C \right)^{
        1/(d_{1} - 1)}
         \left\{ \sinh \left[
             (d_{1} - 1) \sqrt{\epsilon / \mid \theta_{1} \mid} \mid \tau \mid
                       \right]
         \right\}^{-1/(d_{1} - 1)}
\ee
where $C$ is defined by (\ref{17}).
The interval $( - \infty, - 0 ]$ describes the expanding universe and
the interval $[+ 0,  + \infty )$ describes the contracting universe.
It is not difficult to obtain for the modulus of $\varphi$ (taking $c_{n}
= 0$ in (\ref{11})) the formula
\be{44}
\mid \varphi \mid = \left[ \mid \nu_{n} \mid \sqrt{\mid
                        \theta_{1} \mid /\epsilon}~/(d_{1} - 1)
                 \right]
                 \mbox{arsinh} \left(
                   \sqrt{\epsilon/\mid \theta_{1}
\mid}/Ca_{1}^{d_{1}-1}
                 \right)
\ee

It is convenient to rewrite the metric in conformal time $\eta$ which
is connected with harmonic time $\tau$ by
\be{45}
 \sinh \left[ (d_{1} - 1) \sqrt{\epsilon/\mid \theta_{1} \mid} (- \tau)
 \right] = \left\{\sinh [(d_{1} - 1)\eta ]
           \right\}^{-1}
\ee
Then
\be{46}
g = a_{1}^{2}(\eta)\left[ -d\eta\otimes d\eta + g_{(1)} \right]
         + a^{2}_{(0)2}g_{(2)} + \dots + a^{2}_{(0)n} g_{(n)}
\ee
and the scale factor $a_{1}$ as a function of the conformal time is
given by
\be{47}
a_{1}(\eta) = \left( \sqrt{\epsilon/\mid \theta_{1} \mid}/C \right)^{1/(d_{1}
         - 1)}
         \left\{ \sinh \left[
             (d_{1} - 1) \mid \eta \mid
         \right] \right\}^{1/(d_{1} - 1)}
\ee
In synchronous time $t$ the metric takes the form
\be{48}
g = - dt\otimes dt + a_{1}^{2}(t) g_{(1)} +
\sum_{i = 2}^{n} a_{(0)i}^{2} g_{(i)}
\ee
where the scale factor $a_{1}$ and the time coordinate $t$ are
connected by
\be{49}
t = \int \frac{a_{1}^{d_{1} - 1}da_{1}}{\sqrt{\epsilon/\left(|\theta_{1} |
C^{2}\right) + a_{1}^{2(d_{1} - 1)} }} + const
\ee
If $a_{1} \ll \left[ \sqrt{\epsilon/\mid \theta_{1} \mid}/C
\right]^{1/(d_{1} - 1)} $  the scale factor $a_{1}$ has the asymptotic
behaviour $a_{1} \sim \mid t \mid^{1/d_{1}}$. It corresponds to the
open Friedmann universe filled with radiation for $d_{1} = 2$ and
filled with ultrastiff matter for $d_{1} = 3$. If $a_{1} \gg \left[
 \sqrt{\epsilon/\mid \theta_{1} \mid}/C \right]^{1/(d_{1} - 1)} $
the scale factor $a_{1}$ has the asymptotic behaviour corresponding to
a Milne universe $a_{1} \sim \mid t \mid$ for all $d_{1}$. Therefore,
in the case of spontaneous compactification the topology of the
universe asymptotically tends to $M^{d_{1} + 1} \times T^{d_{2}}
\times \dots \times T^{d_{n}}$, where $M^{d_{1} + 1}$ denotes the
$d_{1}+1$ dimensional observable Milne universe (for $d_{1} = 3$ this
is the Minkowky space-time) and the $T^{d_{i}}$ represent the frozen
(unobservable) internal spaces which are $d_{i}$ dimensional tori or
other compact spaces of constant zero curvature (later on we shall
call this topology $M\times T$ topology).

For $d_{1} = 2$ the integral (\ref{49}) can be expressed by elementary
functions
\be{50}
a_{1} = \left[ \sqrt{\epsilon/\mid \theta_{1} \mid}/C  \right]
        \left[ \left( C \sqrt{\mid \theta_{1} \mid / \epsilon} \mid t \mid +
             1 \right)^{2} - 1 \right]^{1/2}
\ee
but for $d_{1} > 2$ we get elliptic integrals.

%%%%%%%%%%%%%%%%%%%%%%%%%%%%%%%%%%%%%%%%%%%%%%%%%%%%%%%%%%%%%%%%%%%%%%%
\subsection{The case $\epsilon < 0$}
%%%%%%%%%%%%%%%%%%%%%%%%%%%%%%%%%%%%%%%%%%%%%%%%%%%%%%%%%%%%%%%%%%%%%%%
%%%%%%%%%%%%%%%%%%%%%%%%%%%%%%%%%%%%%%%%%%%%%%%%%%%%%%%%%%%%%%%%%%%%%%%

In this case, as we can see from equation (\ref{12}), there are
classically allowed and forbidden regions. Classically  forbidden
regions are  usually  treated as regions with Euclidean signature
and classically allowed ones as regions with Lorentzian signature.
As we shall see in the next chapter, on the quantum level there are
tunneling solutions which describe processes with metric signature
alteration \cite{18},
%%%%%%%%%%%%%%%%%%
% 5
%%%%%%%%%%%%%%%%%%%5
for example, universe nucleation from
''nothing''\cite{26}.
%%%%%%%%%%%%%%%%%%%%%%%%%%
%%%  9
%%%%%%%%%%%%%%%%%%%%%%%%%5
If $\epsilon < 0$ at least some of the $\nu_{i}$ should be imaginary.
This leads to complex metric and scalar field. As stressed above it is
possible to have real metric and imaginary scalar field in Lorentzian
region if we demand that $\nu_{i} ~(i = 1, \dots, n-1) $ are real and
$\nu_{n}$ is imaginary and the $\nu_{i}$ should satisfy the condition
(\ref{14}).

Let us consider the equations (\ref{10}), (\ref{12}) in the classically
allowed region $\exp(2qv^{0}) \ge \epsilon/\theta_{1}$. The solutions of
the equations of motion in harmonic time gauge are
\be{51}
\exp(qv^{0}) =\frac{\sqrt{\epsilon/\theta_{1}}}{cos[(d_{1} -
1)\sqrt{\epsilon/\theta_{1}}\tau]}, \qquad \mid \tau \mid \le
\frac{\pi}{2(d_1-1)}\sqrt{\frac{\theta_1}{\epsilon}} \equiv \tau_{1}
\ee
where the constant of integration $\tau_1$ is fixed by a proper choice
of the origin of the time coordinate $\tau$, and
\ba{52}
v^{i} & = &\nu_{i} \tau + c_{i}, \qquad i = 1, \dots, n - 1 \nonumber \\
\varphi & = & \nu_{n} \tau + c_{n}
\ea
The solution in the classically forbidden region can be found with the
help of an analytic continuation $\tau \rightarrow  - i\tau$ in the
expressions (\ref{51}), (\ref{52}).
\newline
As in the case $\epsilon > 0$ we consider here two special cases.

%%%%%%%%%%%%%%%%%%%%%%%%%%%%%%%%%%%%%%%%%%%%%%%%5
\subsubsection{The 2-component universe. Dynamical compactification}
%%%%%%%%%%%%%%%%%%%%%%%%%%%%%%%%%%%%%%%%%%%%%%%%%%%%%%%%%%%%%%%%%%%%%

In this case the scale factors as functions of the harmonic time
coordinate read
\ba{53}
a_{1}^{d_{1} - 1} & = & \frac{a_{(0)1}^{d_{1} -
1}}{\exp\left[\sqrt{\frac{d_{2}(d_{1} - 1)}{D - 2}}~\nu_{1}\tau
\right]\cos \left[\sqrt{\frac{d_{1} - 1}{d_{1}}(\mid \nu_{2}\mid^{2} -
\nu_{1}^{2})}~\tau \right]}, \nonumber \\
a_{2}^{d_{2}} & = & a_{(0)2}^{d_{2}} \exp\left[\sqrt{\frac{d_{2}(d_{1} - 1)}
{D - 2}}~\nu_{1}\tau \right]
\ea
where $\mid \tau \mid \le \tau_{1}$ and the constants $a_{(0)1}$ and
$a_{(0)2}$ are defined by eqns. (\ref{25}), (\ref{26}) with the
substitution $\nu_1^2+\nu_2^2 \rightarrow |\nu_2|^2-\nu_1^2$.
Corresponding to (\ref{11}) the scalar field is given by $\varphi = i
\mid \nu_{2} \mid \tau + c_{2}$. It can be seen from (\ref{53}) that
there are different types of development of the scale factors in
dependence from the sign of $\nu_{1}$. The qualitative picture is
shown  in fig. 11 and fig. 12. Fig. 11 corresponds to $\nu_{1} > 0$
where the scale factor $a_{2}$ increases monotonically from the
minimal value at $\tau = - \tau_{1}$ to the maximal value at $\tau =
\tau_{1}$, while  $a_{1}$ decreases from $+ \infty$ at $\tau = -
\tau_{1}$ down to ${a_{1}}_{\mbox{\scriptsize min}}$ and after that
tends to $+ \infty$ at $\tau = \tau_{1}$. If the value
${a_{2}}_{\mbox{\scriptsize max}}$ does not exceed $L_{Pl}$ too much
for $\tau \rightarrow \tau_{1} $ dynamical compactification arises.
Fig. 12 corresponds to $\nu_{1} < 0$. Here, $a_{2}$ monotonically
decreases from ${a_{2}}_{\mbox{\scriptsize max}}$  at $\tau = -
\tau_{1}$  to ${a_{2}}_{\mbox{\scriptsize min}}$ at $\tau =  \tau_{1}$
and $a_{1}$ once more decreases from $+ \infty$ at $\tau = - \tau_{1}$
down to ${a_{1}}_{\mbox{\scriptsize min}}$ and after that tends to $+
\infty$ at $\tau = \tau_{1}$. In this case the region $0 \le \tau \le
\tau_{1}$ gives an example of dynamical compactification for $\tau
\rightarrow \tau_1$.

In contrast to the case $\epsilon > 0 $ for  $\epsilon < 0 $ we are
not able to give the explicit expression for the scale factors as
functions of the synchronous time coordinate if more than one scale
factors have a dynamical behaviour.

As it was shown in \cite{18}
%%%%%%%%%%%%%%%%%%%%%%%%%%%%%%%%%%
%  5
%%%%%%%%%%%%%%%%%%%%%%%%%%%%%%%%%%
the quantum solutions with fixed value of $\epsilon < 0$ describe
transitions from the classically forbidden Euclidean region to the
classically allowed Lorentzian one (s. next section). The universe is
created by quantum tunneling with the scale $e^{qv_{0}} =
\sqrt{\epsilon/\theta_{1}}$ what cooresponds to the time $\tau = 0$ in
eqns. (\ref{51}) - (\ref{53}). After that the evolution of the
universe is described by the classical equations (\ref{51}) -
(\ref{53}). Figs. 11, 12 show that we have two possible pictures for
the evolution of the universe. Due to the first possibility (fig. 11)
the universe is created by quantum tunnelling with the scale factors
$a_{1} = a_{(0)1}$ and $a_{2} = a_{(0)2}$. After that the space
$M_{1}$ shrinks to $a_{1} = {a_{1}}_{\mbox{\scriptsize min}}$ and then
increases to $+ \infty$. The space $M_{2}$ expands monotonically to
$a_{2} = {a_{2}}_{\mbox{\scriptsize max}}$. The second possibility
(fig. 12) describes a universe created with $a_{1} = a_{(0)1}$ and
$a_{2} = a_{(0)2}$. Here $M_{1}$ expands monotonically to $+ \infty$
while $M_{2}$ contracts to $a_{2} = {a_{2}}_{\mbox{\scriptsize min}}$.
%The latter case represents an %attractive picture of a possible
evolution of the universe if %$a_{(0)1} \sim L_{PL}$ and $a_{(0)2}
\geq L_{PL}$. These parameter %values solve the horizon problem in the
external universe $M_1$, and %if ${a_{2}}_{\mbox{\scriptsize max}}
\leq L_{PL}$ they solve the %problem of compactification.

In conclusion of this section we would like to mention the existence of
the special case of a $n ~(n > 2)$ component model which may be
reduced to the two-component model considered above (s. the Appendix).

%%%%%%%%%%%%%%%%%%%%%%%%%%%%%%%%%%%%%%%%%%%%%%%%%%%
\subsubsection{The n- component universe. Spontaneous
compactification}
%%%%%%%%%%%%%%%%%%%%%%%%%%%%%%%%%%%%%%%%%%%%%%%%%%%

{}From the considerations above we conclude that spontaneous
compactification corresponds to the case $\nu_{i} = 0 ~(i = 1, \dots, n
- 1), \nu_{n} \ne 0$. The condition $\epsilon < 0$ restricts
spontaneous compactification to the presense  of a purely imaginary
scalar field ($\nu_{n}$ imaginary). All the scale factors
$a_{i}=a_{(0)i} ~(i = 2, \dots, n)$ are frozen and
$a_{1} = e^{\beta^{1}}$ has a dynamical
behaviour, only. From (\ref{51}), (\ref{52})
and the transformation (\ref{7}) we find in the Lorentzian region the
scale factor $a_{1}$ as a function  of the harmonic time coordinate
\be{54}
a_{1}(\tau) = \left( \sqrt{\epsilon/ \theta_{1}}/C \right)^{1/(d_{1}
         - 1)}
         \left\{ \cos \left[
             (d_{1} - 1) \sqrt{\epsilon/ \theta_{1}} \tau
                       \right]
         \right\}^{-1/(d_{1} - 1)}
\ee
where $C$ is defined by (\ref{17}). In the interval $\left[ -
\frac{\pi}{2(d_{1} - 1)} \sqrt{ \theta_{1}/\epsilon }, 0 \right]$ the
universe contracts from infinity to the classical turning point where
$a_{1}(\tau = 0) = \left( \sqrt{\epsilon/ \theta_{1}}/C \right)^{1/
(d_{1} - 1)}$ and after that in the interval $\left[ 0,
\frac{\pi}{2(d_{1} - 1)} \sqrt{ \theta_{1}/\epsilon } \right]$ it
expands to infinity again.

It is not difficult to obtain the absolute value of the scalar field
in dependence of the scale factor $a_{1}$ (taking $c_n=0$ in (\ref{52}) )
\be{55}
\mid \varphi \mid = \left[
\mid \nu_{n} \mid  \frac{\sqrt{ \theta_{1}/\epsilon } }
{d_{1} - 1}\right] \arccos \left[ \frac{\sqrt{\epsilon/
\theta_{1}}}{ C a_{1}^{d_{1} - 1}} \right]
\ee
This formula shows that $\mid \varphi \mid $ has it's minimum at the
turning point and tends asymptotically to $\frac{\pi \mid \nu_n \mid }
{2(d_{1} - 1)} \sqrt{ \theta_{1}/\epsilon }$ when $a_{1} \rightarrow
\infty$. The space-time metric takes in harmonic time gauge the form
(\ref{18}), where $a_1(\tau)$ is defined by  (\ref{54}).

The harmonic time $\tau$ and the conformal time $\eta$ are connected
to each other by the expression
\be{56}
\cos \left[(d_{1} - 1) \sqrt{\epsilon/\theta_{1}} \tau \right] =
\left\{ \cosh\left[ (d_{1} - 1)\eta  \right]\right\}^{- 1},
\qquad - \infty < \eta < + \infty
\ee
and the metric in conformal time takes the form (\ref{46}) where
$a_{1}$ depemds on $\eta$ as
\be{57}
a_{1}(\eta) =  \left[ \sqrt{\epsilon/\theta_{1}}/C
\right]^{1/(d_{1} - 1)}
\left\{\cosh\left[ (d_{1} - 1)\eta  \right]\right\}^{ 1/(d_{1} - 1)},
\qquad - \infty < \eta < + \infty
\ee

In order to investigate the asymptotic behaviour of the scale
factor at large times let us consider synchronous coordinates
whith the metric (\ref{48}). We find for the time dependence of the
scale factor $a_{1}$
\be{58}
t = \int \frac{a_{1}^{d_{1} - 1}da_{1}}{\sqrt{a_{1}^{2(d_{1} - 1)} -
\epsilon/(\theta_{1}C^{2}) }} + const
\ee
Then asymptotically, when  $a_{1} \gg \left[
\sqrt{\epsilon/\theta_{1}}/C
\right]^{1/(d_{1}-1)}$, we have $a_{1} \sim | t |$. Thus, the
universe behaves asymptotically like a Milne Universe with respect to the
scale factor $a_{1}$. Therefore, as in the case of spontaneous
compactification for $\epsilon > 0$ (paragraph 3.2.2) the topology of
the universe asymptotically tends to $M\times T$.

In the particular case $d_{1} = 2$ we have from (\ref{58})
\be{59}
a_{1}^{2}(t) = t^{2} + \epsilon/(\theta_{1}C^{2}), \qquad - \infty < t <
+ \infty
\ee
For $d_{1} > 2$ this integral can be expressed with the help of
elliptic integrals. For example, in the case $d_{1} = 3$ we have
\be{60}
t = \left[ \frac{1}{C}\sqrt{\frac{\epsilon}{\theta_{1}}}
        \right]^{1/2} \left\{
               \frac{1}{\sqrt{2}}{\em F}\left( \Psi, \frac{\sqrt{2}}{2} \right)
               - \sqrt{2} {\em E}\left( \Psi, \frac{\sqrt{2}}{2} \right)
               \right\} + \frac{1}{a_{1}}\left[a_{1}^{4} -
            \frac{\epsilon}{\theta_{1}C^{2}} \right]^{1/2}
\ee
where
\be{61}
\Psi = \arccos \left[ \left( \sqrt{\frac{\epsilon}{\theta_{1}}}/C
\right)^{1/2} /a_{1}  \right]
\ee
and {\em F} and {\em E} are the elliptic integrals of  first and
second kind repectively.

The classical expressions for the  Euclidean region $e^{qv^{0}} <
(\epsilon/\theta_{1})^{1/2}$ can be found by analytic continuation of
the formulas obtained here. Then the point
\be{62}
a_{1} = \left[ \sqrt{\epsilon/\theta_{1}}/C \right]^{1/(d_{1} - 1)}
\ee
is the classical turning point. The nucleation of the universe can be
considered as the quantum tunneling process through the potential
barrier \cite{26}.
%%%%%%%%%%%%%%%%%%%%%%%%%%%%%%
%%%%  9
%%%%%%%%%%%%%%%%%%%%%%%%%%%%%%
The universe is nucleated with finite size
$a_{1} = \left[ \sqrt{\epsilon/\theta_{1}}/C \right]^{1/(d_{1} - 1)}$
and zero speed $(da_{1}/dt = 0)$ and its further evolution is
described by classical formulas (\ref{54}), (\ref{57}), and
(\ref{58}).
%%%%%%%%%%%%%%%%%%%%%%%%%%%%%%%%%%%%%%%%%%%%%%%%%%%%%%%%%%%%%%%%%%%%%

\section{SOLUTIONS TO THE QUANTIZED MODEL}

%%%%%%%%%%%%%%%%%%%%%%%%%%%%%%%%%%%%%%%%%%%%%%%%%%%%%%%%%%%%%%%%%%%55

At the quantum level the constraint equation (\ref{9}) turns over to
the Wheeler-De Witt equation (WDW). The WDW equation is covariant with
respect to minisuperspace coordinate transformations and can be written in
the harmonic time gauge in the following form \cite{14}
%%%%%%%%%%%%%%%%%%%%%%
%%%   1
%%%%%%%%%%%%%%%%%%%%%5
\be{63}
\left( - \frac{\partial^{2}}{\partial {v^{0}}^{2}} +
         \frac{\partial^{2}}{\partial {v^{1}}^{2}}
       + \dots +
         \frac{\partial^{2}}{\partial {v^{n - 1}}^{2}}
       + \frac{\partial^{2}}{\partial {\varphi}^{2}}
       - \mid \theta_{1} \mid e^{2qv^{0}}
 \right) \Psi  = 0
\ee
It is easy to obtain solutions of the WDW equation (\ref{63}) by
separation of variables
\be{64}
\Psi = \Psi_{0}(v^{0}) \dots \Psi_{n -1}(v^{n - 1})\Psi_{n}(\varphi)
\ee
where
\be{65}
\Psi_{i}(v^{i}) = e^{i\nu_{i}v^{i}}, i = 1, \dots, n - 1; \qquad
\Psi_{n}(\varphi) =  e^{i\nu_{n}\varphi}
\ee
and $\Psi_{0}$ satisfies the equation
\be{66}
\left(
- \frac{d^{2}}{{dv_{0}}^{2}} - \mid \theta_{1} \mid e^{2qv^{0}}
\right) \Psi_{0} = \epsilon \Psi_{0}
\ee
Here, $\epsilon$ and the arbitrary numbers $\nu_{i}$ are related to
each other by
\be{67}
\epsilon = \sum_{i = 1}^{n} \nu_{i}^2
\ee
The solutions of equation (\ref{66}) are
\be{68}
C^{(m)}_{i\sqrt{\epsilon}/q} \left[ \left( \sqrt{\mid \theta_{1}
\mid}/q \right) e^{qv^{0}} \right]
\ee
where $C^{(m)}$ denotes the Bessel function of the first $(m = 1)$,
second ($m = 2)$ or third $(m = 3)$ kind. It was shown in
\cite{14}
%%%%%%%%%%%%%%%%%%%%%%
%%%  2
%%%%%%%%%%%%%%%%%%%%%%
that $\epsilon$ can be interpreted as an energy. From this point of
view the states with $\epsilon = 0$ are treated as ground states.
The general solution can be written in the form
\be{69}
\Psi = \sum_{m = 1}^{3} \int d^{n}\nu B^{(m)}(\nu)
C^{(m)}_{i\sqrt{\epsilon}/q} \prod_{j = 1}^{n}  e^{i\nu_{j}v^{j}}
\ee
where the $B^{(m)}(\nu)$ are arbitrary coefficients depending on the quantum
numbers $\nu_{i}$.

The solutions (\ref{64}) are eigenstates of the quantum mechanical
operators
\be{70}
\Pi_{v^{i}} = - \frac{i}{N}\frac{\partial}{\partial v^{i}},
i = 1, \dots, n - 1, \qquad
\Pi_{\varphi} = - \frac{i}{N}\frac{\partial}{\partial \varphi}
\ee
with the eigenvalues $\nu_{n}/N$, where  we have $N = 1$ for the
Lorentzian space-time and $N = i$ for the Euclidean one. The classical
equations corresponding to the states (\ref{64}) are \cite{14}
\be{71}
\dot v^{i} = N^{-1} \nu_{i}, \qquad i = 1, \dots, n - 1, \qquad \dot\varphi =
N^{-1}\nu_{n}
\ee
where the dot denotes the differentiation with respect to harmonic
time $\tau$. Evidently, equations (\ref{71}) coinside with (\ref{10}),
(\ref{11}). Thus, for the classical equations corresponding to (\ref{64})
the constants of integration $\nu_{i}$ in (\ref{11}) should coinside with the
quantum numbers $\nu_{i}$ in (\ref{65}).

Let us consider the wave function (\ref{64}) in more detail. In the
same way as in the classical case we distinguish three special cases:
$\epsilon = 0, > 0, < 0$.

%%%%%%%%%%%%%%%%%%%%%%%%%%%%%%%%%%%%%%%%%%%%%%%%%%%%%%%%%%%%%%%%%%%%%

\subsection{The case $\epsilon = 0$}

%%%%%%%%%%%%%%%%%%%%%%%%%%%%%%%%%%%%%%%%%%%%%%%%%%%%%%%%%%%%%%%%%%%%
%%%%%%%%%%%%%%%%%%%%%%%%%%%%%%%%%%%%%%%%%%%%%%%%%%%%%%%%%%%%%%%%%%%%

This case was considered earlier
in \cite{14}, where the state with $\epsilon = 0$ was treated as the
ground state.
For all $\nu_i$ to be real the condition $\epsilon = 0$ leads to the
demand $\nu_i=0 ~(i = 1, \dots, n)$.
As shown above, this
corresponds in the classical limit to an universe with spontaneous
compactification and the $M\times T$ topology. Once more, this
describes the product of an $d_1+1$
dimensional Milne Universe and static $d_i$ dimensional tori or other
compact spaces of constant zero curvature.  The wave function
\be{72}
\Psi_{0}^{(0)} = J_{0} \left[ \left( \sqrt{\mid \theta_{1}
\mid}/q \right) e^{qv^{0}} \right] = J_{0} \left[ \left( \sqrt{\mid \theta_{1}
\mid}/q \right) a_{1}^{d_{1} - 1}a_{(0)2}^{d_{2}} \dots a_{(0)n}^{d_{n}}
              \right]
\ee
is related to the Hartle-Hawking boundary conditions \cite{27}
and describes the superposition of expanding and contracting
universes \cite{14}.
The wave function
\be{73}
\Psi_{0}^{(2)} = H_0^{(2)} \left[ \left( \sqrt{\mid \theta_{1}
\mid}/q \right) e^{qv^{0}} \right]
\ee
describes the expanding universe and satisfies the Vilenkin boundary
conditions \cite{14,28}.
The wave function
\be{74}
\Psi_{0}^{(1)} = H_0^{(1)} \left[ \left( \sqrt{\mid \theta_{1}
\mid}/q \right) e^{qv^{0}} \right]
\ee
describes the contracting universe.  Here  $J_{0}$ and
$H_{0}^{(1,2)}$ are the Bessel functions of first and third kind
respectively.

The Hartle-Hawking ground state wave function (\ref{72}) is
non-singular. The vacuum wave functions (\ref{73}) and (\ref{74}) go
to infinity as $\ln a_{1}$ when $a_{1} \rightarrow 0$.

%%%%%%%%%%%%%%%%%%%%%%%%%%%%%%%%%%%%%%%%%%%%%%%%%%%%%
\subsection{The case $\epsilon > 0$}
%%%%%%%%%%%%%%%%%%%%%%%%%%%%%%%%%%%%%%%%%%%%%%%%%%%%%
%%%%%%%%%%%%%%%%%%%%%%%%%%%%%%%%%%%%%%%%%%%%%%%%%%%%%

Let us assume all $\nu_{i}$ to be real. This corresponds to real
metric and scalar field in the Lorentzian region.

The excited states (\ref{64}) can be written in the form
\be{75}
\Psi_{\nu_{1}, \dots, \nu_{n}}^{(0)} = J_{ik} \left[ \left(
\sqrt{\mid \theta_{1} \mid}/q \right) e^{qv^{0}} \right]
\prod_{j = 1}^{n} e^{i\nu_{j}(v^{j} - c^{j})}
\ee
\be{76}
\Psi_{\nu_{1}, \dots, \nu_{n}}^{(1,2)} = H_{ik}^{(1,2)} \left[ \left(
\sqrt{\mid \theta_{1} \mid}/q \right) e^{qv^{0}} \right]
\prod_{j = 1}^{n} e^{i\nu_{j}(v^{j} - c^{j})}
\ee
where the energy $\epsilon = \sum_{1}^{n} \nu_{i}^{2} \equiv
q^{2}k^{2}, - \infty < k < + \infty, ~\varphi \equiv v^{n}$, and the
translation invariance of the WDW equation (\ref{63}) was used for the
$v^{i} ~(i = 1, \dots, n)$.

All these wave functions oscillate an infinite number of times when
the spacial geometry degenerates $(a_{i} \rightarrow 0)$. This
singular behaviour reflects the initial and final singularities of the
classical solutions.

As shown above the solutions of the classical Lorentzian equations
have the following asymptotic behaviour. When $\tau \rightarrow 0$ we
have $\mid t \mid \rightarrow \infty$ and $v^{i} \rightarrow c^{i}$ so
that $a_{i} \rightarrow a_{(0)i}$ $(i = 2, \dots, n)$ and
$a_{1} \sim \mid t \mid $. This corresponds asymptotically to a
freezing of the internal dimensions, and the limit of dynamical
compactification is spontaneous compactification.
Therefore, spontaneous compactification acts as an attractor solution
for solutions with dynamical compactification.
The corresponding
limit for the wave functions (\ref{75}), (\ref{76}) may be achieved
for $v^0 \rightarrow \infty, v^i \rightarrow c^i (i = 1, \dots, n)$.
In this case the asymptotic bahaviour of the
solutions (\ref{75}), (\ref{76}) is given by
\be{77}
\Psi^{(0)} = \cos \left[ \left(
\sqrt{\mid \theta_{1} \mid}/q \right) a_{1}^{d_{1} - 1}a_{(0)2}^{d_{2}}
 \dots a_{(0)n}^{d_{n}} \right]
\ee
\be{78}
\Psi^{(j)} = \exp \left[(- 1)^{j+1} i  \left(
\sqrt{\mid \theta_{1} \mid}/q \right) a_{1}^{d_{1} - 1}a_{(0)2}^{d_{2}}
 \dots a_{(0)n}^{d_{n}} \right]
\ee
where $j = 1,2$ and (\ref{77}), (\ref{78}) describe  in
the classical  limit asymptotically a universe with the $M \times T$
topology.

Evidently, the solutions (\ref{75}, \ref{76}) are not the unique solutions
of equation (\ref{63}), other wave functions can also be found.
Some of these wave functions may be free of the above mentioned
singularities. For the one component model with scalar field this was
analysed in \cite{29}.
In analogy to the papers \cite{15,16},
where the case $\theta > 0$ was considered,  we can give here an example
of such a solution. We write the wave function (\ref{76}) in the form
\be{79}
\Psi^{(j)}_{k, \gamma_{1}, \dots, \gamma_{n - 1}} =
    \exp \left[i k \sum_{l = 1}^{n}q(v^{l} - c^{l}) \Gamma_{l} \right]
    H^{(j)}_{ik} \left[\left(\frac{\sqrt{\mid \theta_{1} \mid}}{q}\right)
    e^{qv^{0}} \right]
\ee
where the quantum numbers $\nu_{i}$ are written in the form
\be{80}
\nu_{i} = kq\Gamma_{i}
\ee
and the $n$ dimensional unit vector $\Gamma_{i}$ is defined by
\be{81}
\Gamma_{i} = \left(
               \begin{array}{llll}
                 {\cos \gamma_{1}} & & &  \\
                 {\sin \gamma_{1}} & {\cos \gamma_{2}} & &  \\
                 {\sin \gamma_{1}} & {\sin \gamma_{2}} & {\cos \gamma_{3}} & \\
                 \vdots          &                 &                 & \\
                 {\sin \gamma_{1}} & \dots  & {\sin \gamma_{n - 2}} &
                           {\cos
                       \gamma_{n - 1}}  \\
                 {\sin \gamma_{1}} & \dots  & {\sin \gamma_{n - 2}} & { \sin
                       \gamma_{n - 1}}  \\
               \end{array}
             \right)
\ee
Using the integral transformation \cite{30}
\be{82}
\Psi^{(j)}_{\lambda, \gamma_{1}, \dots, \gamma_{n - 1}} =
\int_{- \infty}^{+ \infty} dk  C_{k}^{(j)}(\lambda)
\Psi^{(j)}_{k, \gamma_{1}, \dots, \gamma_{n - 1}}
\ee
with
\be{83}
C_{k}^{(j)}(\lambda) = \frac{i}{2} (- 1)^{j + 1} \exp \left[(-1)^{j}\frac{\pi
k}{2}\right]\exp(- ik\lambda)
\ee
where $-\infty < \lambda < \infty$, we get
\be{84}
\Psi^{(j)}_{\lambda, \gamma_{1}, \dots, \gamma_{n - 1}} =
\exp \left\{
     (-1)^{j+1} i \left(\frac{\sqrt{\mid \theta_{1} \mid}}{q}  \right)
       e^{qv^{0}} \cosh\left[\sum_{i=1}^{n}q(v^{i} - c^{i})\Gamma_{i}
- \lambda  \right]
     \right\}
\ee
This wave function has the same asymptotic behaviour at $v^{i}
\rightarrow c^{i}~(i = 1, \dots, n),~v^{0} \rightarrow + \infty,$ as
(\ref{79}). It corresponds asymptotically to the contracting $(j=1)$ and
expandig $(j=2)$ $(d_i+1)$ dimensional Milne universe.
The wave function (\ref{84}) can already not descibe a state of fixed
energy $\epsilon$.

%%%%%%%%%%%%%%%%%%%%%%%%%%%%%%%%%%%%%%%%%%%%%%%%%%%%%%%%%%%%%%%%%%%%%%%
\subsection{The case $\epsilon < 0$}

%%%%%%%%%%%%%%%%%%%%%%%%%%%%%%%%%%%%%%%%%%%%%%%%%%%%%%%%%%%%%%%%%%%%%%%
%%%%%%%%%%%%%%%%%%%%%%%%%%%%%%%%%%%%%%%%%%%%%%%%%%%%%%%%%%%%%%%%%%%%%%%

In this case the wave functions (\ref{64}) can be written in the form
\be{85}
\Psi_{\nu_{1}, \dots, \nu_{n}}^{(0)} = J_{k} \left[ \left(
\frac{\sqrt{\mid \theta_{1} \mid}}{q} \right) e^{qv^{0}} \right]
\prod_{j = 1}^{n} e^{i\nu_{j}(v^{j}-\tilde c^j)}
\ee
\be{86}
\Psi_{\nu_{1}, \dots, \nu_{n}}^{(1,2)} = H_{k}^{(1,2)} \left[ \left(
\frac{\sqrt{\mid \theta_{1} \mid}}{q} \right) e^{qv^{0}} \right]
\prod_{j = 1}^{n} e^{i\nu_{j}(v^{j}-\tilde c^j)}
\ee
where $\varphi \equiv v^{n}, j = 1,2, \epsilon \equiv - q^{2}k^{2}, -
\infty < k < \infty$ and for $v^j ~(j=1, \dots, n)$ the translation
invariance of the WDW equation (\ref{63}) was used.

As a consequence of the condition $\epsilon < 0$ the quantum numbers
$\nu_{i}$ (or part of them) became imaginary. Then, in the
classically allowed region we would get a complex metric.
In order to avoid this we
demand in analogy to the classical case all $\nu_{i} ~(i = 1, \dots, n
- 1)$ to be real and $\nu_{n}$ to be imaginary and the condition
(\ref{14}) to be valid. Consequently, with these conditions the metric
of the classically accessible region is a real Lorentz metric, while
the scalar field in this region is imaginary.

It can be seen from equation (\ref{66}) (the quantum analog of
(\ref{12})) that in the case $\epsilon < 0$ there exist classically
accessible as well as forbidden regions. The solutions (\ref{85}),
(\ref{86}) are solutions with fixed energy $\epsilon$
and describes transitions between the classically allowed and forbidden
regions due to tunneling processes. As a consequence, it becomes
possible to analyze processes with changes of the metric signature
\cite{18}.
For instance, the solution $\Psi^{(2)}$ in (\ref{86}) describes
quantum tunneling through a potential barrier and is usually interpreted
as creation of the universe from ''nothing'' \cite{14,18,26}. In this way
for $\nu_{i} = 0 ~(i = 1, \dots, n - 1)$ a $n$ component universe with
spontaneous compactification will be created, while for $\nu_1 \neq 0,
\nu_{i} = 0 ~(i = 2, \dots, n - 1)$ the
creation of a $n$-component universe with
dynamical compactification is described. The solution $\Psi^{(1)}$
describes the opposite process, the transition  into ''nothing'' as
the final stage of the evolution of the universe.

In analogy to the case $\epsilon > 0$ we have also solutions to
equation (\ref{63}), which already do not describe wave functions for
fixed energy. For instance, with an integral transformation of
$\Psi^{(0)}$ we find \cite{30}
\ba{87}
\Psi^{(0)}_{\lambda, \gamma_{1}, \dots, \gamma_{n - 1}} & = &
\int_{- \infty}^{+ \infty} dk  C_{k}(\lambda)
\Psi^{(0)}_{k, \gamma_{1}, \dots, \gamma_{n - 1}}  \nonumber \\
& = & \exp \left\{
      i \left(\frac{\sqrt{\mid \theta_{1} \mid}}{q}  \right)
       e^{qv^{0}} \sin\left[\sum_{i=1}^{n - 1}q(v^{i} -
     \tilde c^{i})\Gamma_{i}\sinh\gamma_{n - 1} + \right. \right. \\
& & + \left.\left. qi(v^{n} - \tilde c^{n})\cosh\gamma_{n - 1}
+ \lambda  \right] \right\}   \nonumber
\ea
where $C_k(\lambda) = \exp(ik\lambda)$, $\lambda$ has a continuous
spectrum of width $2\pi$ and the wave function (\ref{85})
$\Psi^{(0)}_{\nu_1 \dots \nu_n}$ was refwritten in the form
\ba{88}
\Psi^{(0)}_{k, \gamma_{1}, \dots, \gamma_{n - 1}} & = &
     \exp \left\{
      ik  \left[\sum_{i=1}^{n - 1}q(v^{i} -
              \tilde c^{i})\Gamma_{i}\right.\sinh\gamma_{n - 1}\right.
                 +  \nonumber \\
& & + \left.\left. qi(v^{n} - \tilde c^{n})\cosh\gamma_{n - 1}\right]
     \right\}\times J_{k} \left[\frac{\sqrt{\mid \theta_{1} \mid}}{q}
       e^{qv^{0}} \right]
\ea
where we used for the quantum numbers $\nu_i$ the representation
\be{89}
\nu_{i} = kq\Gamma_{i}\sinh\gamma_{n - 1}, \qquad i = 1, \dots, n - 1
\ee
\be{90}
\nu_{n} = ikq \cosh\gamma_{n - 1}
\ee
Here $\Gamma_{i}$ represents a $(n - 1)$ dimensional unit vector given
by a formula similar to (\ref{81}) and the arbitrary constants $\tilde
c^{i}$ can be taken best as
the classical limits for the $v^{i} ~(i = 1, \dots, n)$ in equation
(\ref{52}) for $\mid \tau \mid  \rightarrow \tau_{1}$, i.e. $\tilde
c^i=\nu^i\tau_1 + c^i$ or $\tilde c^i= -\nu^i\tau_1 + c^i$. In this limit
the $v^{i} ~(i = 1, \dots, n)$ tend to their maximal (or minimal) fixed
values and the corresponding scale factors $a_{i} ~(i = 2, \dots, n)$
are frozen out and $a_{1} \sim \mid t \mid ~\rightarrow \infty$. So it
can be seen that in the limit $v^{i} \rightarrow \tilde c^i ~(i = 1, \dots,
n), v^{0} \rightarrow \infty$, the wave
functions  $\Psi^{(0)}_{k, \dots} $ and $\Psi^{(0)}_{\lambda, \dots}$ have
the same asymptotic behaviour and describe asymptotically in the classical
limit an universe with the same $M\times T$ topology like in the case
$\epsilon > 0$.

%%%%%%%%%%%%%%%%%%%%%%%%%%%%%%%%%%%%%%%%%%%%%%%%%%%%%%%%%%%%%%%%%

\section{CONCLUSIONS}

%%%%%%%%%%%%%%%%%%%%%%%%%%%%%%%%%%%%%%%%%%%%%%%%%%%%%%%%%%%%%%%%%
We investigated multidimensional cosmological models (MCM) with $n ~(n
>1)$ Einstein spaces for the case these Einstein spaces were of
constant curvature. The integrable case was considered where only one
of these spaces was of negative constant curvature, while all others
were assumed to be Ricci flat. As a matter source we introduced a
massless minimally coupled homogeneous scalar field. The main
attention was paid to the problem of campactification of extra
dimensions. The non Ricci flat space $M_1$ was considered as our external
space while all the other spaces with zero curvature described
internal spaces. But our general solutions do not exclude the
possiblility that one of the Ricci flat spaces may play the role of
our external space.

The problem of compactification for the model with positive curvature
of the non Ricci flat space was considered in detail in our previous
paper \cite{10} where solutions with dynamical and spontaneous
compactification were found. In the case of positive curvature of the non
Ricci flat space the parameter $\epsilon$ defined by formula
(\ref{13}) plays the role of energy \cite{14} and  should be positive in
the Lorentzian region. For the model with negative curvature of the
non Ricci flat space considered in the  present paper we have in the
Lorentzian region for this parameter $\epsilon \ge 0$ as well as
$\epsilon < 0$. This feature leads to a more complex picture than in
the former case. For all values of the parameter $\epsilon$ solutions
with dynamical and spontaneous compactification were found.
It is not difficult to see that the solution with spontaneous
compactification in the case $\epsilon = 0$ is an attractor solution
for all kinds of solutions with $\epsilon > 0$ as well as $\epsilon <
0$. In the limit of large goemetry all solutions tend to $a_1 \sim |t|
\rightarrow \infty$ for the scale factor of the space $M_1$ while all
other scale factors and the scalar field become freezed (see figures 7
- 10, 13). Thus,
asymptotically all solutions describe the universe with
the topology
$M^{d_1+1}\times T^{d_2}\times \dots \times T^{d_n}$ where $M^{d_1+1}$
is the $(d_1+1)$- dimensional Milne universe and the $T^{d_1}$ are
$d_i$-dimensional frozen tori or other compact spaces of constant zero
curvature.
Solutions
to the quantum Wheeler-DeWitt equation were obtained also. In the
case $\epsilon < 0$ some of them describe the process of tunneling
{}from the Euclidean region to the Lorentzian one, often called the birth
of the universe from ''nothing'' \cite{28}. For all values of
$\epsilon$ these wave functions asymptotically describe in the
classical limit an universe with the topology
$M^{d_1+1}\times T^{d_2}\times \dots \times T^{d_n}$.

\bigskip

\noindent
{\bf Acknowledgment}\\
The work was sponsored by the WIP project
016659/p. One of us (A. Z.) was supported by DAAD and by DFG grant
436UKR-17/7/93.
A. Z. also thanks Prof. Kleinert and the Freie Universit\"at Berlin for
their hospitality.

\newpage

%%%%%%%%%%%%%%%%%%%%%%%%%%%%%%%%%%%%%%%%%%%%%%%%

\section{APPENDIX}
\renewcommand{\theequation}{A.\arabic{equation}}
\setcounter{equation}{0}
%%%%%%%%%%%%%%%%%%%%%%%%%%%%%%%%%%%%%%%%%%%%%%%%
%%%%%%%%%%%%%%%%%%%%%%%%%%%%%%%%%%%%%%%%%%%%%%%%
It is easy to see that there exist a special $n$-component case $(n >
2)$ which can be reduced to the two component universe with dynamical
compactification considered above for $\epsilon > 0$ as well as for
$\epsilon < 0$.
This exceptional case corresponds to the special
choice of the integration constants in equation (\ref{11})
\ba{A1}
\nu_{1} & \ne & 0  \nonumber   \\
\nu_{2} &  = & \dots = \nu_{n - 1} = 0
\ea
Than it follows from the coordinate transformation (\ref{7}) that
\be{A2}
a_{i} = e^{B_{i}}a_{2}, \qquad i = 3, \dots, n,
\ee
where $B_{i}$ is an arbitrary constant. Therefore, all factor spaces
$M_{i} ~(i= 3, \dots, n)$ have identical dynamical behaviour, the same
as $M_{2}$ in the two component universe. Further, the equations
(\ref{23}) - (42) and (\ref{53}) remain the same and define the
dynamics of the factor spaces $M_{1}$ and $M_{2}$ with the only
difference that we have to make the change $d_{2} \rightarrow \sum_{2}
= \sum_{i=2}^{n} d_{i}$.

\newpage
\noindent
{\bf Figures}

\vspace*{7cm}

\noindent
{\bf Figure~1} The dynamical behaviour of the scale factors $a_1$ and
$a_2$ in harmonic time for $\epsilon > 0$ in the case 1. if $\nu_1 >
0$.

\vspace{9cm}
{\bf Figure~2} The dynamical behaviour of the scale factors $a_1$ and
$a_2$ in harmonic time for $\epsilon > 0$ in the case 1. if $\nu_{1} < 0$.
\newpage

%\noindent
%{\bf Figures}

\vspace*{7cm}

\noindent
{\bf Figure~3} The dynamical behaviour of the scale factors $a_1$ and
$a_2$ in harmonic time for $\epsilon > 0$ in the case 2. if $\nu_1 >
0$.

\vspace{9cm}
{\bf Figure~4} The dynamical behaviour of the scale factors $a_1$ and
$a_2$ in harmonic time for $\epsilon > 0$ in the case 2. if $\nu_1 <
0$.

\newpage

%\noindent
%{\bf Figures}

\vspace*{7cm}

\noindent
{\bf Figure~5} The dynamical behaviour of the scale factors $a_1$ and
$a_2$ in harmonic time for $\epsilon > 0$ in the case 3. if $\nu_1 >
0$.

\vspace{9cm}
{\bf Figure~6} The dynamical behaviour of the scale factors $a_1$ and
$a_2$ in harmonic time for $\epsilon > 0$ in the case 3. if $\nu_1 <
0$.

\newpage

%\noindent
%{\bf Figures}

\vspace*{7cm}

\noindent
{\bf Figure~7} The dynamical behaviour of the scale factors $a_1$ and
$a_2$ in synchronous time for $\epsilon > 0$ in the case 3. if $d_1 = 3$
(formulas (\ref{37}), (38). The line $a_1 = |t|$ is the attractor for
the scale factor $a_1$.

\vspace{9cm}
{\bf Figure~8} The dynamical behaviour of the scale factors $a_1$ and
$a_2$ in synchronous time for $\epsilon > 0$ in the case 3. if $d_1 = 3$
(formulas (\ref{40}), (41). The line $a_1 = |t|$ is the attractor for
the scale factor $a_1$.

\newpage

%\noindent
%{\bf Figures}

\vspace*{7cm}

\noindent
{\bf Figure~9} The case 3. $(d_{1} = 3)$. The behaviour of the scalar
field for equation (39) (with positive sign).

\vspace{9cm}
{\bf Figure~10} The case 3. $(d_{1} = 3)$. The behaviour of the scalar
field for equation (42) (with positive sign).

\newpage

%\noindent
%{\bf Figures}

\vspace*{7cm}

\noindent
{\bf Figure~11} The qualitative behaviour of the scale factors $a_1$
and $a_2$ in harmonic time for $\epsilon < 0$ (equation
 (\ref{53}) for $\nu_{1} > 0)$.

\vspace{9cm}
{\bf Figure~12} The qualitative behaviour of the scale factors $a_1$
and $a_2$ in harmonic time for $\epsilon < 0$ (equation
 (\ref{53}) for $\nu_{1} <0)$.
It can be seen that in the case $a_{2 min} \sim L_{Pl}$
for  $\tau \rightarrow \tau_{1}$  dynamical compactification takes place.

\newpage
\vspace*{10truecm}

{\bf Figure~13} For solutions with spontaneou compactification of all
internal spaces and negative curvature of the external space it is
shown how the solution $a_1 = |t|$ for $\epsilon = 0$ acts as an
attractor. The curves are drawn for the simple case $d_1 = 2$. Here we
have $a_1 = \left[ t^2 +
\frac{\epsilon}{\theta_1}\frac{1}{C^2}\right]^{1/2}$ for $\epsilon <
0$ and $a_1 = \sqrt{\frac{\epsilon}{|\theta_1|}}\frac{1}{C}\left[
\left( C \sqrt{\frac{|\theta_1|}{\epsilon}}|t| + 1 \right)^2 - 1
\right]^{1/2}$ for $\epsilon > 0$. For all dynamical solutions
$\epsilon {> \atop <} 0$ we have asymptotically $a_1 \sim |t|
\rightarrow + \infty, ~a_2, \dots, a_n \rightarrow $ const, $ ~\varphi
\rightarrow$ const  and, therefore, the solution $a_1 = |t|, ~\epsilon
= 0$ is an attractor, too (see e.g. figures 7 - 10).

%*****************************************************************
%*****************************************************************

%%%%%%%%%%%%%%%%%%%%%%%%%%%%%%%%%%%%%%%%%%%%%%%%%%%%%%%%%%%%%%%%
%%%%%%%%%%%%%%%%%%%%%%%%%%%%%%%%%%%%%%%%%%%%%%%%%%%%%%%%%%%%%%%%
%\end{enumerate}


\begin{thebibliography}{99}
\bibitem{22}  Originally, the Milne universe is the four-dimensional $(d =
       3)$ hyperbolic space-time with the scale factor  $a(t) = t$
       \cite{23,24}.
       In our multidimensional model (\ref{1}) we call the space-time
       a Milne universe in the more general case when all scale
       factors are freezed with the exception of one of them, for example
       $a_{1}$ which depends on synchronous time as $a_{1}(t) = t$.
       and the space $M_{1}$ is the space of negative constant curvature with
       the arbitrary dimension $d_{1} > 1$.
\end{thebibliography}
\end{document}